\documentclass[referee]{mn2e}
\usepackage{graphicx}

\title[H-alpha survey of nearby dwarf galaxies]
{H-alpha survey of nearby dwarf galaxies}

\author[S. S. Kaisin et al.]
{S. S. Kaisin$^{1}$\thanks{E-mail:skai@sao.ru},
I.~D.~Karachentsev$^{1}$,
S.~Ravindranath$^{2}$\\
$^{1}$Special Astrophysical Observatory, Russian Academy of Sciences,
    N.Arkhyz, Russia \\
$^{2}$Inter-University Centre for Astronomy and Astrophysics, Pune, India}
 \begin{document}
\pagerange{\pageref{firstpage}--\pageref{lastpage}} \pubyear{2008}

\def\LaTeX{L\kern-.36em\raise.3ex\hbox{a}\kern-.15em
   T\kern-.1667em\lower.7ex\hbox{E}\kern-.125emX}

\newtheorem{theorem}{Theorem}[section]

\label{firstpage}

\maketitle

\begin{abstract}

  We present the $H\alpha$ imaging data and flux measurements for 30  dwarf
galaxies in the Local volume. The $H\alpha$ fluxes
are used to derive the galaxy star formation rate, SFR. The sample of
observed galaxies is characterized by the following parameters: the median
distance of 7.5 Mpc, the median blue absolute magnitude of $-14.8^m$,
and median SFR of --2.0 dex. Two dSph members of the Local Group: Cetus
and Leo~IV do not show signs of star formation on the rate of --5.4 dex
and --7.0 dex, respectively. The BCD galaxy ESO~553-46 has one of the
highest specific SFR among the Local volume galaxies.
\end{abstract}

\begin{keywords}
dwarf galaxies: star formation
\end{keywords}

\section{Introduction}

Since 2004, a survey program to map $H\alpha$ emission in galaxies
belonging to the Local volume (LV) with distance $D < 10$ Mpc has been
underway using the 6-m telescope at the Special Astrophysical Observatory
of Russian Academy of Sciences (SAO RAS).
This project is approaching completeness, and the basic
results have been published in a series of papers:
Karachentsev et al.(2005), Kaisin \& Karachentsev (2006, 2008),
Kaisin et al. (2007, 2011), and Karachentsev \& Kaisin (2007, 2010).
In parallel to our survey, similar observations of nearby galaxies
in the $H\alpha$ line were performed also in Cambridge and Arizona
universities, which review is published by Kennicutt et al. (2008). At
present, the degree of completeness of the $H\alpha$ survey of the LV objects
exceeds 90\% in the Northern hemisphere, but the observations
in the Southern hemisphere have not been so successful so far. Essential
input to the $H\alpha$ survey of northern LV galaxies was introduced by
observations with the 6-meter BTA telescope of SAO RAS. For a continuation
of the survey into the southern sky we used facilities of the 2-meter
telescope of Girawali observatory (IUCAA) situated at the geographic
latitude of $+18^{\circ}$. In this paper we present results from the observing
runs during 2009-2010, which have yeilded $H\alpha$ fluxes and SFR for 30 nearby
dwarf galaxies.

\section{Observations and image processing}

Observations of the nearby dwarf galaxies were performed using the
2-m telescope at the IUCAA Girawali Observatory, located near Pune,
India. The average seeing was 1.5$^{\prime\prime}$  during most of the observing
period. Images of the galaxies were recorded with the IUCAA Faint
Object Spectrograph and Camera instrument, equipped with a EEV $2K\times 2K$ CCD which
provides a 10.5$^{\prime}$ field of view with a resolution of 0.31$^{\prime\prime}$ per pixel.
The images were obtained via an interference $H\alpha$ filter with an
effective wavelength 6563\AA \ and bandwidth FWHM = 84\AA, and also with the
standard wide-band $R$-filter to subtract a continuum. The typical exposure
time was 1200 sec in the $H\alpha$ line and 600 sec in the continuum.
Since the range of radial velocities in our sample is small, we used one
and the same filters for all the observed objects.

 The ESO-MIDAS astronomy data reduction package was used to process our
obtained data. To confirm the quality of the results, comparisons between
different nights and standard stars were done.
 The observational data were processed in the standard way that included:
bias subtracting, flat-fielding by twilight flats, removing cosmic rays,
and the sky background subtracting. The next operation was to align
the emission-line image and continuum ones. Here, PSF matching was done
to match the resolutions of the $H\alpha$ - image to continuum images.
Then the images in the continuum were normalized to
$H\alpha$ images using 10 -- 20 stars and subtracted.
The measured $H\alpha$ fluxes of the galaxies were calibrated
with spectrophotometric standard stars from Hamuy et al. (1992, 1994)
obtained in the same night. The internal errors for the $H\alpha$ flux
measurements were typically about 0.10~dex. The major contribution to the
error was from variable atmospheric conditions.
 The conditions of our observations are given in the column 11 Table 1,
colon means that the sky was not photometric.
We did not correct
$H\alpha$ fluxes for the emission of the [NII] doublet, which is likely
to be small, $\sim$0.05~dex, in the case of dwarf galaxies (Lee et al. 2009).

 Images of the galaxies we have observed are shown as a mosaic in Fig.1.
The left and right images of each galaxy correspond to the  $H\alpha$ plus
continuum and in the $H\alpha$ minus continuum. The angular
scale and north and east directions are indicated by the horizontal bars
and by the arrows.

 The main characteristics of the observed galaxies are listed in Table 1.
The table columns contain: (1) --- galaxy name; (2) --- equatorial coordinates
for epoch J2000.0; (3) --- morphological type from de Vaucouleurs digital
classification; (4) --- radial velocity relative to the Local Group centroid
(in km s$^{-1}$); (5) --- distance to the galaxy in Mpc with indication of the
method used: ``rgb'' --- via the tip of red giant branch, ``tf'' ---- from the
Tully-Fisher relation, ``mem'' --- from a membership of galaxy in a group,
``h'' --- via the radial velocity for the Hubble parameter $H_0 = 73$ km s$^{-1}$Mpc$^{-1}$;
(6, 7) --- the observed integrated and extinction-corrected $H\alpha$ flux
of the galaxy on a logarithmic scale in units of [erg/cm$^2$/s], here the
Galactic extinction was accounted according to Schlegel et al. (1998) and
the internal extinction via equation (5) from Karachentsev et al. (2004)
assuming the extinction in $H\alpha$ to be 0.538 of extinction in the
$B$--band; (8) --- integrated star formation rate in the galaxy calculated from
canonical relation [SFR] =  0.945$\cdot10^9\cdot F_c(H\alpha) D^2$, where the distance
to it is expressed in Mpc  (Kennicutt, 1998); (9) --- absolute $B$-
magnitude of the galaxy corrected for extinction; (10) ---  logarithm of the
hydrogen mass of the galaxy taken from  HIPASS (Meyer et al. 2004,
Koribalski et al. 2004), (11) --- observation conditions.

\section{Some individual properties of the observed galaxies}

{\em Cetus dSph.}  This dwarf spheroidal galaxy is a peripheric companion
to Andromeda (M~31) being on projected distance of 620 kpc from it. Like
KKR~25 and Apples~I, Cetus is a rare example of isolated dSph system
having not any normal galaxy within 0.5 Mpc. Bouchard et al. (2006) found
an HI--cloud probably associated with the dwarf. $H\alpha$ emission from the
Cetus dSph is not detected by us.  History of star formation in Cetus
was studied by Monelli et al. 2010 basing on deep HST/ACS observations.

{\em UGC 1056.} A blue compact dwarf (BCD), being a companion to the
bright spiral NGC~628. $H\alpha$ and UV emissions are concentrated in its
core.  It was imaged in $H\alpha$ by Kennicutt et al. 2008.

{\em ESO 483-013.} An isolated BCD galaxy with $H\alpha$ emission in the
center. It is mistakenly classified by NED as lenticular type.
 It was imaged in $H\alpha$ by Kennicutt et al. 2008.

{\em UGC 3174 = DDO 34.} A dwarf spiral galaxy with a dozen small
HII- regions.  It was imaged in $H\alpha$ by Kennicutt et al. 2008.

{\em ESO 553-046.} An isolated BCD galaxy with an extremely high star
formation rate per luminosity unit.

{\em UGCA 127 sat.} A dwarf irregular galaxy without radial velocity.
Probably, it is a companion to the spiral UGCA~127, which distance
derived from Tully-Fisher relation was ascribed to the companion itself.
On the western side of the galaxy disc there is an unresolved HII region
apart from the total diffuse $H\alpha$ emission.

{\em WHI B0619-07.} This dwarf spiral galaxy residing in a zone of strong
extinction is an another companion to UGCA~127. A bright star is projected
into its central part.

{\em ESO 490-017.} A nearby dIr galaxy, whose distance was determined via
the tip of red giant branch (Karachentsev et al. 2003).  The HI map
and velocity field of ESO 490-017 were obtained with GMRT by Begum et al. 2008.

{\em CGMW1-260 and IC 2171.} Both the Sdm galaxies have similar radial
velocities and distance estimates from TF-relation.

{\em NGC 2283.} This is a Scd galaxy with a majority of compact HII-
regions. Judging to its radial velocity and TF-distance, NGC~2283 belongs
to a scattered group at low galactic latitude, $\mid b\mid \sim10^{\circ}$, together with
two previous galaxies.

{\em KKSG 9.} A BCD galaxy with a high hydrogen mass-to-luminosity ratio.
Probably, it has an extended HI- envelope that may be verified by
observations in the HI line with aperture synthesis. The distance to
KKSG~9 was estimated from its apparent membership in the NGC~2283 group.

{\em ESO 558-011.} A galaxy of Magellanic (Sm) type with faint HII-regions.

{\em HIPASS J0801-21.} A dwarf irregular galaxy but without signs of
$H\alpha$ emission. Its coordinates in NED are given with an error.

{\em ESO 495-008 and ESO 497-004.} Both the galaxies with low radial
velocities are situated outside the Local volume, judging to their
TF-distance estimates.

{\em HIPASS J0916-23b = ESO 497-035.} A companion to the spiral galaxy
NGC~2835.

{\em ESO 565--003 and 6dFJ0939351-250735.} Isolated dIr galaxies with small
emission knots.

{\em 6dFJ0956376-092911.} A single BCD galaxy undetected in HIPASS.

{\em KKSG 15, MCG-01-26-009, KKSG 17, and UGCA 193.} Probable companions to
the lenticular galaxy NGC~3115.  UGCA 193 was imaged in $H\alpha$ by Kennicutt et al. 2008.

{\em UGC 6145.} This is a dIr companion to the spiral NGC~3521. It shows
faint diffuse $H\alpha$ emission.  The HI map and velocity field of ESO 490-017 were obtained with GMRT by Begum et al. 2008.

{\em Leo IV.} A dwarf spheroidal companion to the Milky Way recently
discovered by Belokurov et al (2007). It does not show any signs of
$H\alpha$ emission.  Its structure and star formation history was
studied by Sand et al. 2010. A probable companion to Leo V (de Jong et al. 2010.

{\em KKSG 33.} A dSph companion to the ``Sombrero''  (NGC~4594) without
signs of emission in HI nor $H\alpha$.

{\em IC 3647 = KDG 180 = VCC 1857.} A dSph galaxy with unreliably measured
radial velocity  from Binggeli et al. 1985.
It is a probable member of the Virgo cluster without
apparent emission in $H\alpha$ and HI.

{\em NGC 4700.} This Sd galaxy has a radial velocity being much more than
the expected one under TF-distance of 6.5 Mpc.

{\em 2MFGC 15085.} One of few galaxies with relatively low radial velocity
situated in front of the Local Void center. A faint diffuse companion
is seen  $\sim2^{\prime}$ NW to it.

\section{Discussion}

 Of the 30 galaxies in Table 1, there are 4 objects: UGC~1056, ESO~483-013,
UGC~3174, and UGCA~193 for which the $H\alpha$ fluxes have been measured by
others. All four estimates were made by Kennicutt et al. (2008). A
comparison of our and their (K08) estimates shows that the average
arithmetic difference  $\langle \log F_{our} - \log F_{K08} \rangle = -0.04\pm0.09$, and the
mean square difference of the logarithm of the fluxes is 0.16. This
indicates a satisfactory agreement of the independent measurements (in spite
of the imperfect weather conditions), although the external flux
error turns out to be about 1.5 times of the internal one.

 The objects observed by us are characterized by a large diversity of
luminosity, morphological types and other parameters. The median distance
to them is 7.5 Mpc, the median absolute blue magnitude is $M_B = -14.8^m$,
and the median star formation rate is log[SFR] = --2.0. Two dwarf spheroidal
member of the Local group: Cetus and Leo~IV do not show any signs of current
star formation on the level of $(10^{-5} - 10^{-7})\cdot M_{\odot}$/year, while the
Sc galaxy NGC~2283 transform its gas into stars with a rate of $\sim 1 M_{\odot}$/year.

The distribution of the observed galaxies versus
their absolute magnitude $M_B$ and SFR is presented in Fig 2.
(squares). The galaxies with only an upper limit for the
$H\alpha$  flux are not indicated in the figure. For comparison, we
also present another 450 galaxies from the Local volume ($ D < 10$),
shown by circles. The SFR data for them are taken
from literature. The line corresponds to a constant SFR per unit luminosity.

Another diagram,  is presented in Fig 3, where the global SFR of
galaxies is compared with their mass of neutral hydrogen $M_{HI}$.
The galaxies with only an upper limit for the SFR or $M_{HI}$ are not indicated.
In this diagram the dashed line corresponds to a
fixed SFR per unit $M_{HI}$ and the solid line traces the known Kennicutt-
Schmidt relatioship  $SFR\propto M_{HI}^{1.5}$.

 As it is seen from the data of columns (8) -- (10) of Table, the star
formation rate of a galaxy correlates with its luminosity, as well as with
the amount of neutral hydrogen. To characterize the evolution status of
a galaxy one can use the ``Past-Future'' diagram \{$P^*, F^*$\}, where the
dimensionless  and distance independent parameters  $P^* = \log([SFR] T/L_B)$ and
$F^* = \log(M_{HI}/[SFR]\cdot T)$
determine the productivity of star formation in the past over the whole
cosmic time T = 13.7 Gyrs, and for how long the star formation
can continue in future with the present gas resource.  Actually, the
$P^*$ parameter is equivalent to the widely used "specific star formation
rate" normalized to the cosmic time T, and the parameter $F^*$ corresponds
to the time scale to exhaust the HI mass (in units of T), also often used
by many authors. The titles of \{$P^*, F^*$\} quadrants indicate conditionally
an evolution state of galaxies in them.

For instance, a galaxy with the constant SFR and the constant mass-to-luminosity
ratio locates in the origin of the \{$P^*, F^*$\}- plane if it is able to
reproduce its observed luminosity over the cosmic time T, and also has the
HI amount to keep the observed SFR during the next T-term.

Most of bulgeless spiral galaxies and irregular dwarfs on the \{$P^*, F^*$\}
diagram are concentrated around the origin \{$P^*=0, F^*$=0\}. Therefore,
bulk of the local late type galaxies reside at a middle-way of their star
forming evolution.

Nevertheless, the diagnostic diagram in Figure 4 exhibits a significant
scatter of galaxies on these parameters. As one can see, the sample
of galaxies in Fig.4 is slightly
elongated along the diagonal $F^*= -P^*$. Such a feature appears
when the star formation in dwarf galaxies (they amounts a majority in
our sample) is characterized by burst activity changing a rate of star
formation over an order of magnitude. In particular, we may consider
the BCD galaxy ESO~553-046 to be just in a stage of star formation burst.
Under the present value of SFR, the galaxy is able to reproduce its total
luminosity during only 1.5 Gyrs, and its gas resource would be sufficient to
keep the present star formation rate on the time scale $ \sim 1/30$ of the
cosmic age.  Many other examples of nearby dwarf galaxies currently
undergoing global starbursts are presented by Lee et al. (2009)

Note that Stinson et al. (2007) simulated evolution of dwarf irregular
galaxies taking into account effects of gas outflows due to the wind from
SNs, and found cyclic bursts of star formation on the scale of 0.3 Gyr
with an amplitude of about (2-3) magnitudes for dwarf systems of very low
mass. On the other hand, as it was shown by Lee et al. (2009), the dwarfs
that are currently experiencing massive global bursts consist of only 6/%
their number, and bursts are responsible for about a quarter of the total
star formation in the dwarf population.

The "Replenished Catalog of Nearby Galaxies" (Karachentsev et al. 2012)
contains 825 galaxies residing in the Local Volume. Among them, there are
about 500 galaxies with measured H-alpha fluxes and about 700 galaxies
having estimates of SFR from FUV- magnitudes measured with GALEX (Martin et
al. 2005). We shall use this extended Local Volume sample to study star
formation properties  of nearby galaxies in more details. \

   {\bf  Acknowledgements}
We thank the referee for valuable comments.
This work was supported by RUS--IND RFBR grant 10--02--92650, RFBR grant
10--02--00123 and  the Ministry of Education and Science of.
the Russian Federation under contract 14.740.11.0901

\clearpage
\begin{table}
\caption{H-alpha fluxes and SFRs for 30 nearby galaxies}
\begin{tabular}{lcrrrlrrrrrc} \\ \hline
Name     &    RA    Dec.    &  $T$ & $V_{LG}$&
\multicolumn{2}{c}{$D$}&
$\log F_{H\alpha}$    &$\log F_{H\alpha}$& $\log SFR$ & $M_B$   &$\log M_{HI}$ & {\bf obs.}        \\
	 &    (2000.0)     &    &     &
\multicolumn{2}{c}{Mpc}    & obs&   corr &  corr &       & &      {\bf cond. }       \\
\hline
 (1)     &        (2)      &(3) & (4) &
\multicolumn{2}{c}{(5)}&
(6)  &  (7)  &   (8) &  (9)  & (10) & {\bf(11)} \\
\hline
Cetus    &  002611.0$-$110240& $-$1 &  26 & 0.78 &rgb &$<-$14.30 &$<-$14.28& $<-$5.53& $-$10.18&  5.87 &{\bf }\\
UGC~1056  &  012847.2$+$164117&  9 & 774 & 7.3   &mem & $-$12.73 & $-$12.67&  $-$1.97& $-$14.82&  7.64 & \\
ESO~483-013&  041241.1$-$230932&  9 & 782 & 7.4  &tf  & $-$12.59 & $-$12.54&  $-$1.83& $-$15.51&  7.62 & \\
UGC~3174  &  044834.5$+$001430&  8 & 609 & 8.3   &h   & $-$12.79 & $-$12.70&  $-$1.89& $-$15.60&  8.38 & \\
ESO~553-046&  052705.7$-$204041&  9 & 370 & 5.1  &h   & $-$11.96 & $-$11.92&  $-$1.53& $-$14.13&  7.22 & \\
UA~127sat &  062054.8$-$083901& 10 &  ---  & 7.5 &mem & $-$13.64 & $-$12.88&  $-$2.26& $-$16.00&   --- &{\bf:}\\
WHI~B0619&  062213.8$-$075016&  8 & 586 & 8.4    &tf  & $-$12.87 & $-$12.28&  $-$1.46& $-$17.73&  8.94 & \\
ESO~490-017&  063756.6$-$255959& 10 & 268 & 4.23 &rgb & $-$12.36 & $-$12.29&  $-$2.07& $-$14.46&  7.55 & \\
CGMW1-260&  063800.1$-$150122&  8 & 542 & 9.9    &tf  & $-$12.67 & $-$12.30&  $-$1.34& $-$18.02&  8.63 &{\bf:} \\
IC~2171  &  064427.3$-$175557&  8 & 565 & 9.9    &tf  & $-$12.53 & $-$12.09&  $-$1.13& $-$16.87&  8.38 & \\
NGC~2283  &  064552.7$-$181237&  6 & 622 &10.0   &tf  & $-$11.49 & $-$11.11&  $-$0.14& $-$18.82&  9.43 & \\
KKSG~9    &  064656.9$-$175629&  9 & 477 &10.0   &mem & $-$13.29 & $-$12.85&  $-$1.89& $-$15.62&  8.63 & \\
ESO~558-011&  070656.8$-$220226&  8 & 496 & 6.7  &tf  & $-$12.83 & $-$12.47&  $-$1.85& $-$16.39&  7.87 & \\
HIPASS~0801-21& 080125.4$-$215951& 10 & 463 &6.4 &h   &$<-$14.38 &$<-$14.17& $<-$3.59& $-$12.79&  7.54 & \\
ESO~495-008&  082100.7$-$234653& 10 & 512 &17.3  &tf  & $-$13.32 & $-$13.21&  $-$1.79& $-$15.62&  8.26 &{\bf:} \\
ESO~497-004&  090303.1$-$234830&  8 & 519 &14.5  &tf  & $-$13.76 & $-$13.54&  $-$2.25& $-$15.46&  8.46 & \\
HIPASS~0916-23b& 091658.0$-$231647& 10& 542&9.3  &mem & $-$13.76 & $-$13.66&  $-$2.75& $-$14.47&  7.42 & \\
ESO~565-003&  092309.9$-$201003& 10 & 549 & 7.6  &h   & $-$13.21 & $-$13.15&  $-$2.42& $-$14.14&  7.33 & \\
6dF~J0939351  &  093935.0$-$250735& 10 & 440&6.0 &h   & $-$13.07 & $-$13.02&  $-$2.49& $-$13.64&  6.97 & \\
6dF~J0956376  &  095637.6$-$092911&  9 & 355&4.9 &h   & $-$13.79 & $-$13.73&  $-$3.38& $-$12.84& $<$6.75&  \\
KKSG~15   &  095510.5$-$061612& 10 & 554 & 9.7   &mem & $-$13.79 & $-$13.75&  $-$2.81& $-$14.97&  7.77 & \\
MCG-01-26-009& 100133.6$-$063130& 10 & 510 & 9.7 &mem & $-$13.30 & $-$13.27&  $-$2.33& $-$13.98&  7.71 & \\
KKSG~17   &  100138.4$-$081456& 10 & 203 & 9.7   &mem & $-$14.12 & $-$14.07&  $-$3.13& $-$14.78&  7.72 & \\
UGCA~193  &  100236.2$-$060043&  7 & 425 & 9.7   &mem & $-$12.94 & $-$12.82&  $-$1.88& $-$15.83&  8.54 & \\
UGC~6145  &  110535.0$-$015149& 10 & 533 &10.7   &mem & $-$14.16 & $-$14.11&  $-$3.08& $-$13.94&  7.97 & \\
Leo~IV   &  113257.0$-$003200& $-$3 & $-$59 & 0.16 &rgb &$<-$14.57 &$<-$14.55& $<-$7.17& $-$ 4.63&  4.23:& \\
KKSG~33   &  124008.9$-$122153& $-$3 &  ---  & 9.3 &mem &$<-$14.40 &$<-$14.36& $<-$3.45& $-$11.54& $<$6.99&  \\
IC~3647   &  124053.1$+$102834& $-$1 & 528:& 7.2 &h   &$<-$14.31 &$<-$14.29& $<-$3.61& $-$14.90& $<$6.44 & \\
NGC~4700  &  124907.6$-$112441&  7 &1196 & 6.5   &tf  & $-$12.54 & $-$12.44&  $-$1.84& $-$16.92&  8.32 & \\
2MFGC~15085& 194311.7$-$065621&  8 &1654 &16.7   &tf  & $-$12.94 & $-$12.60&  $-$1.18& $-$17.19&  9.14 & \\
\hline
\end{tabular}
\end{table}

\onecolumn

  \begin{figure}
 \vbox{\includegraphics{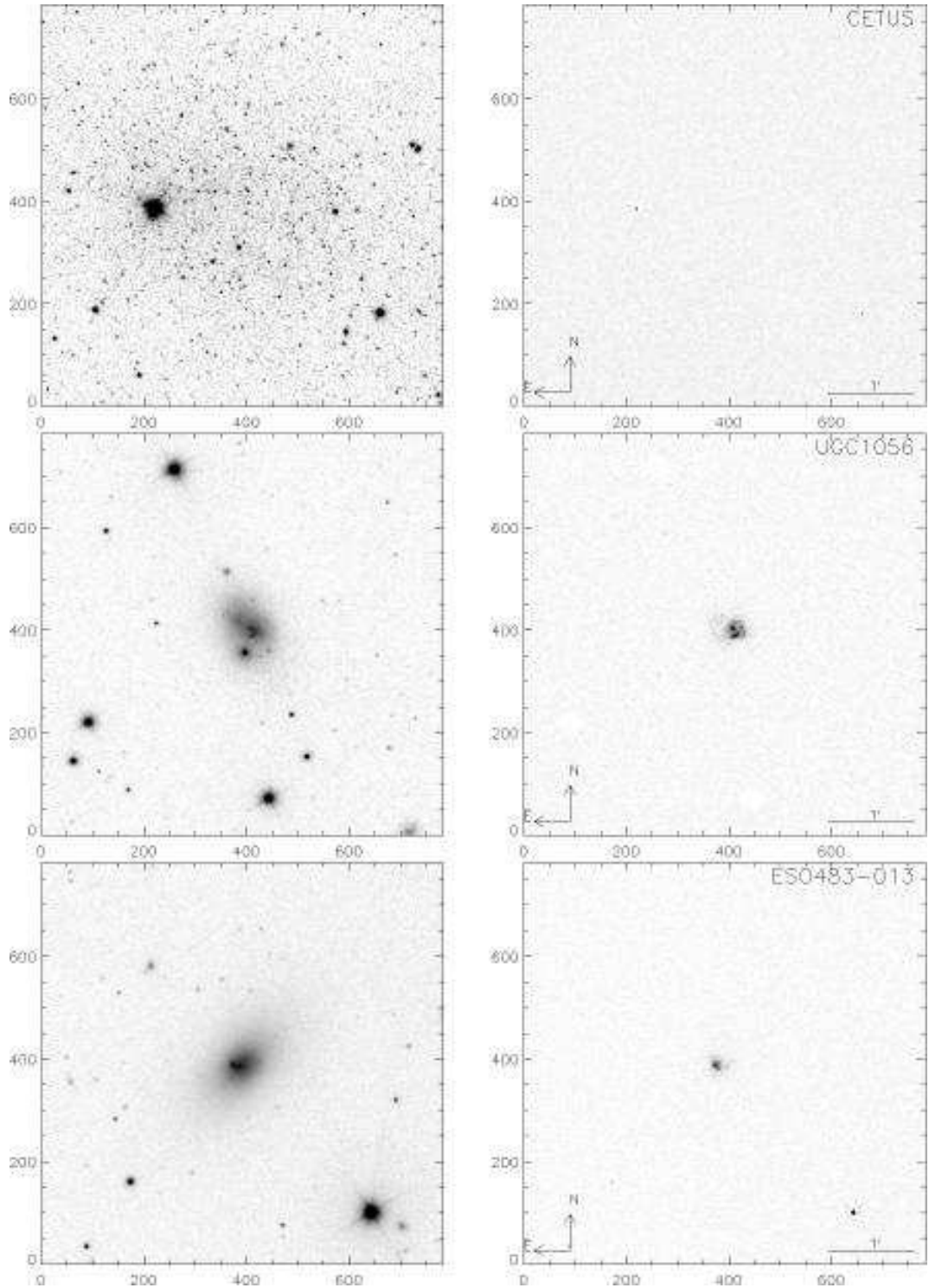}}
\vspace{22cm}
\caption{Mosaic of images of the 30 nearby galaxies. On the left are images
	 in $H\alpha$ + continuum and on the right, in $H\alpha$ with the
	 continuum subtracted. The scale and north and east directions
	 are indicated in corners.}
\end{figure}

  \begin{figure}
\setcounter{figure}{0}
 \vbox{\includegraphics{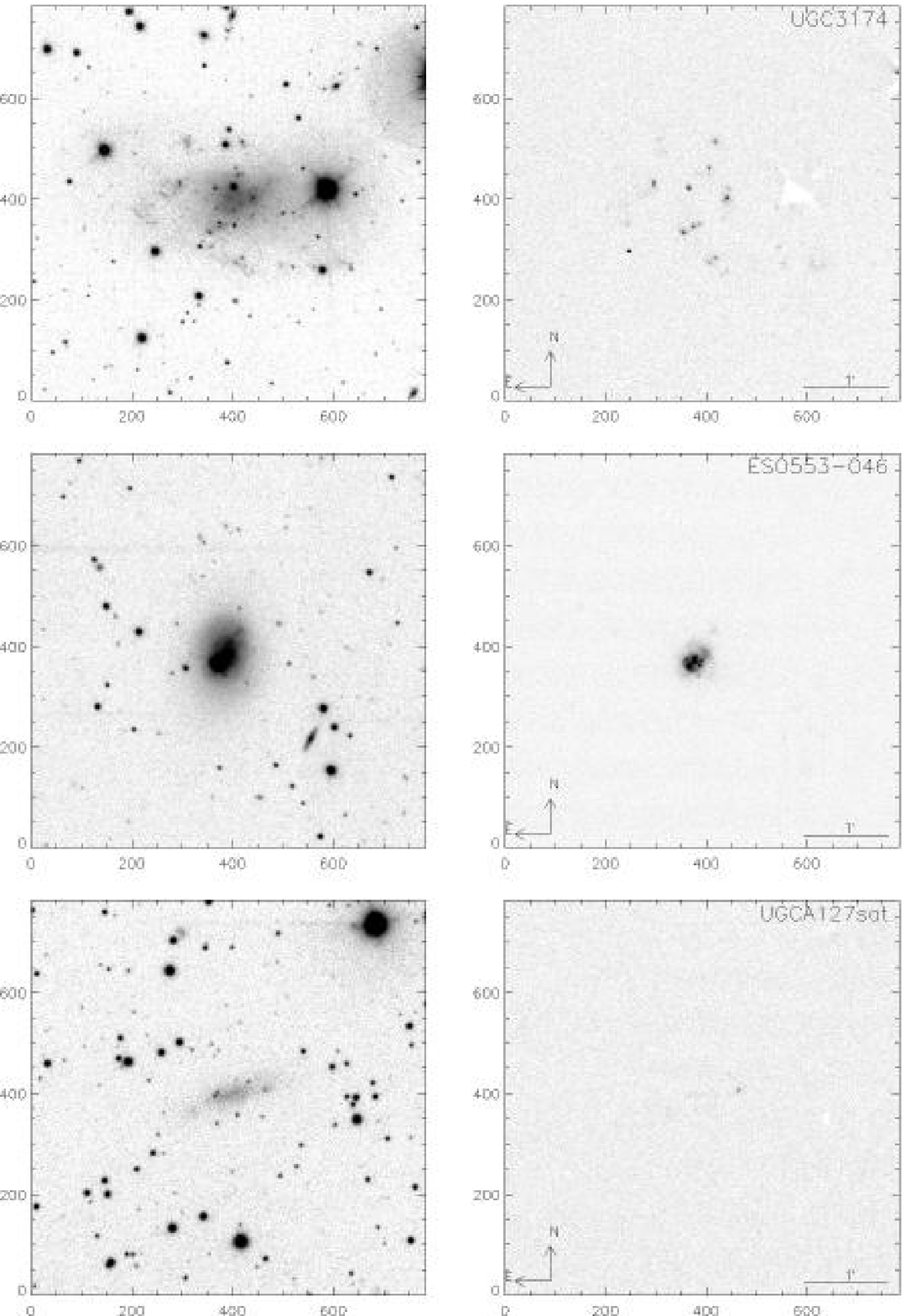}}
\vspace{22cm}
\caption{Continued}
\end{figure}

  \begin{figure}
\setcounter{figure}{0}
 \vbox{\includegraphics{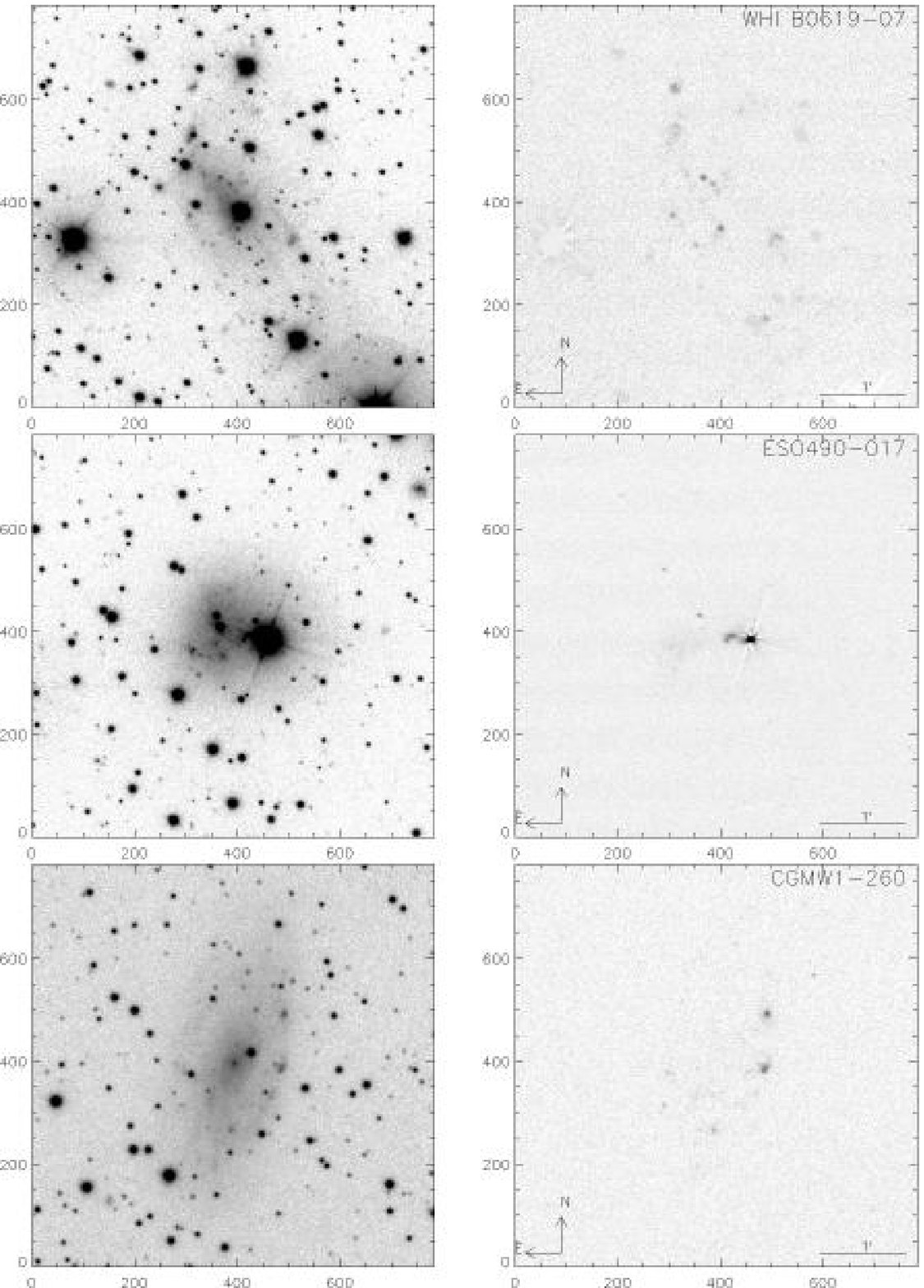}}
\vspace{22cm}
\caption{Continued}
\end{figure}

  \begin{figure}
\setcounter{figure}{0}
 \vbox{\includegraphics{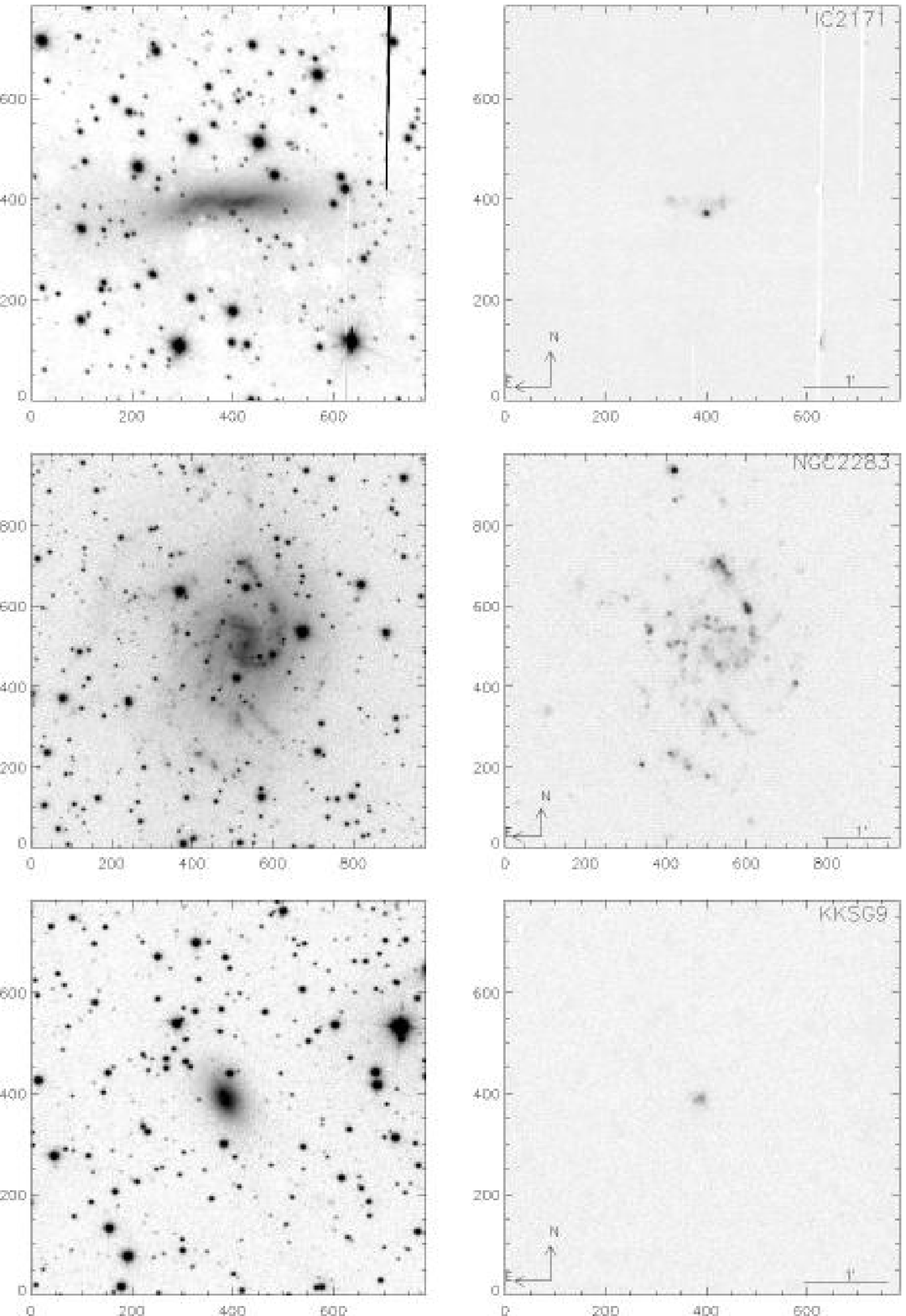}}
\vspace{22cm}
\caption{Continued}
\end{figure}

  \begin{figure}
\setcounter{figure}{0}
 \vbox{\includegraphics{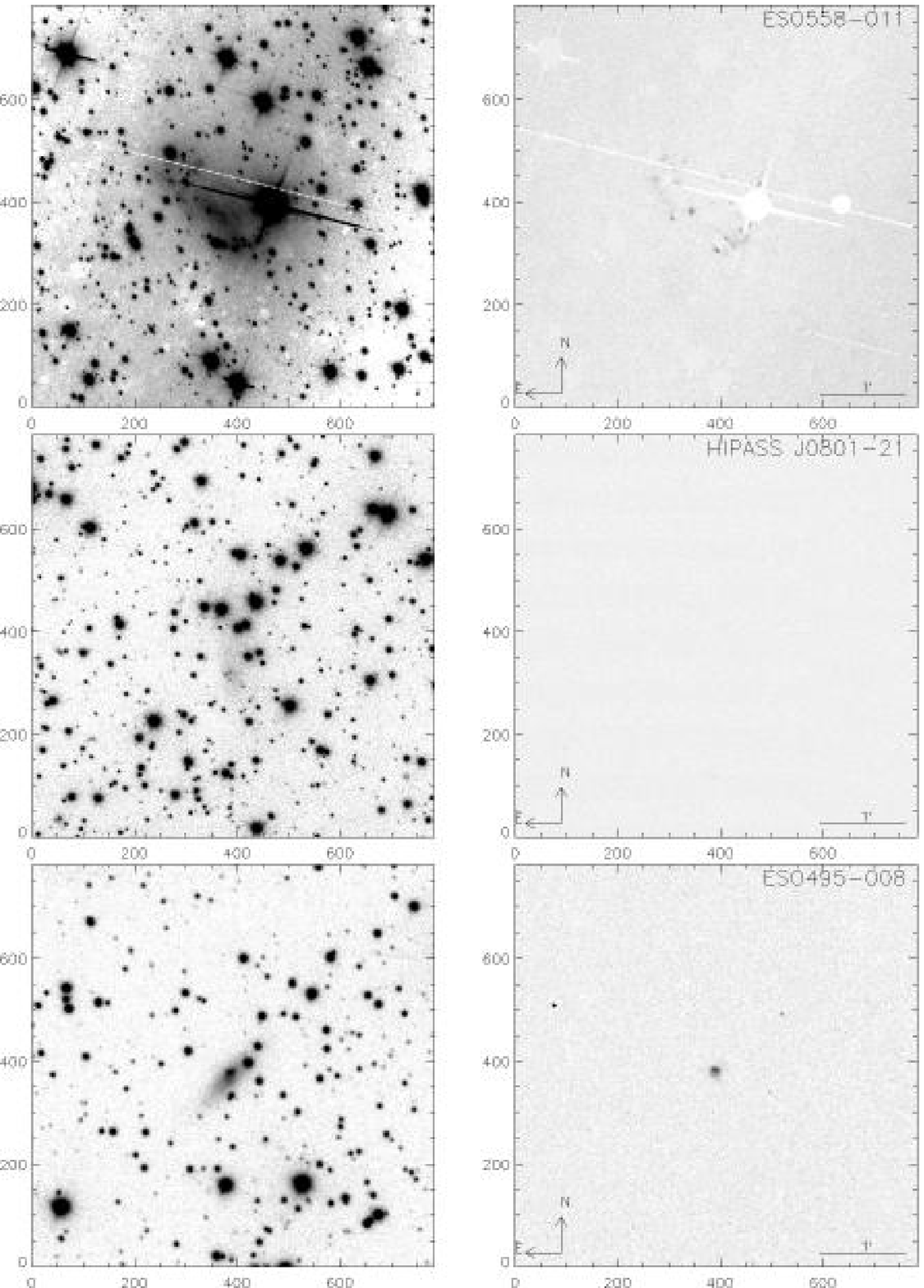}}
\vspace{22cm}
\caption{Continued}
\end{figure}

  \begin{figure}
\setcounter{figure}{0}
 \vbox{\includegraphics{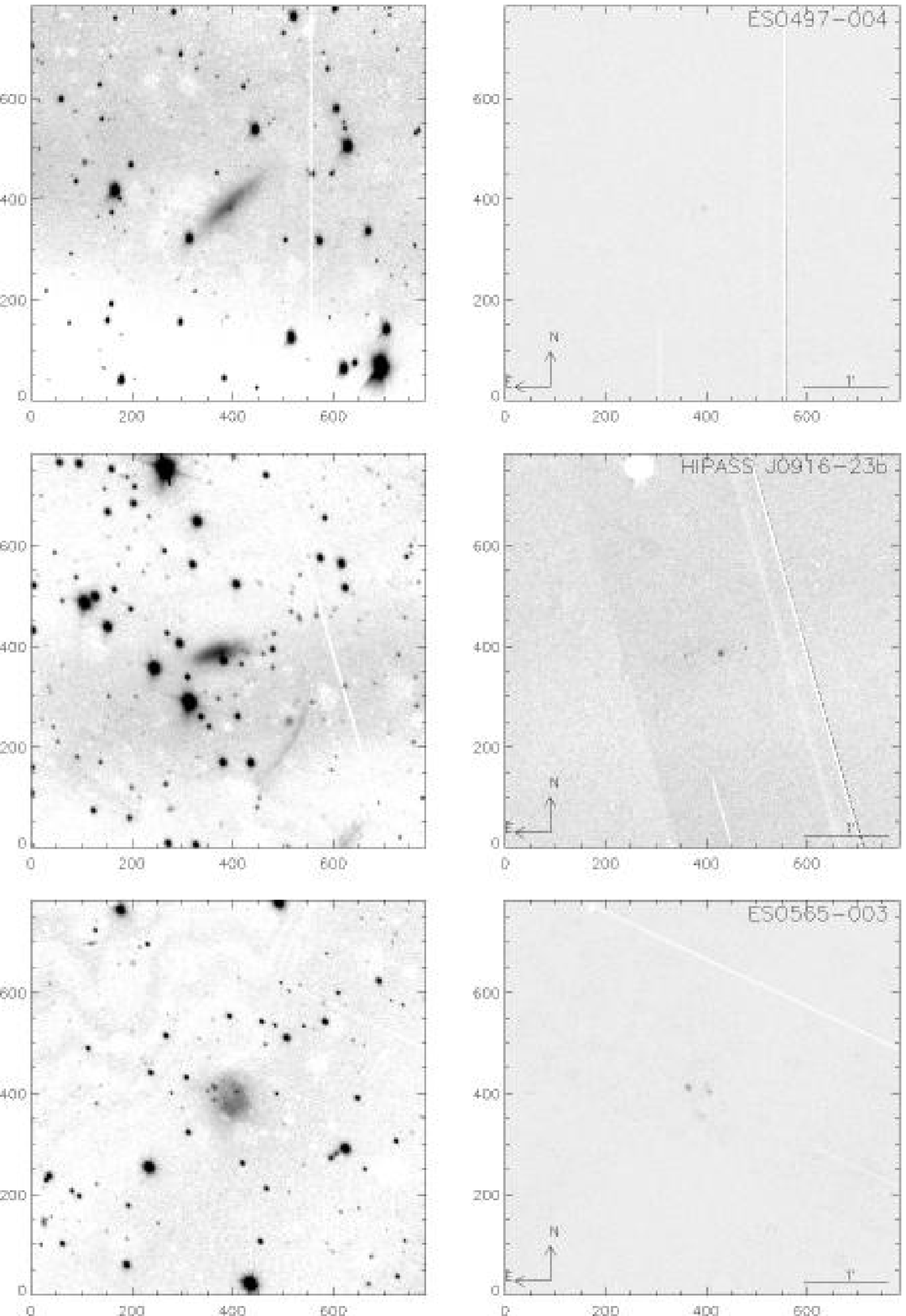}}
\vspace{22cm}
\caption{Continued}
\end{figure}

  \begin{figure}
\setcounter{figure}{0}
 \vbox{\includegraphics{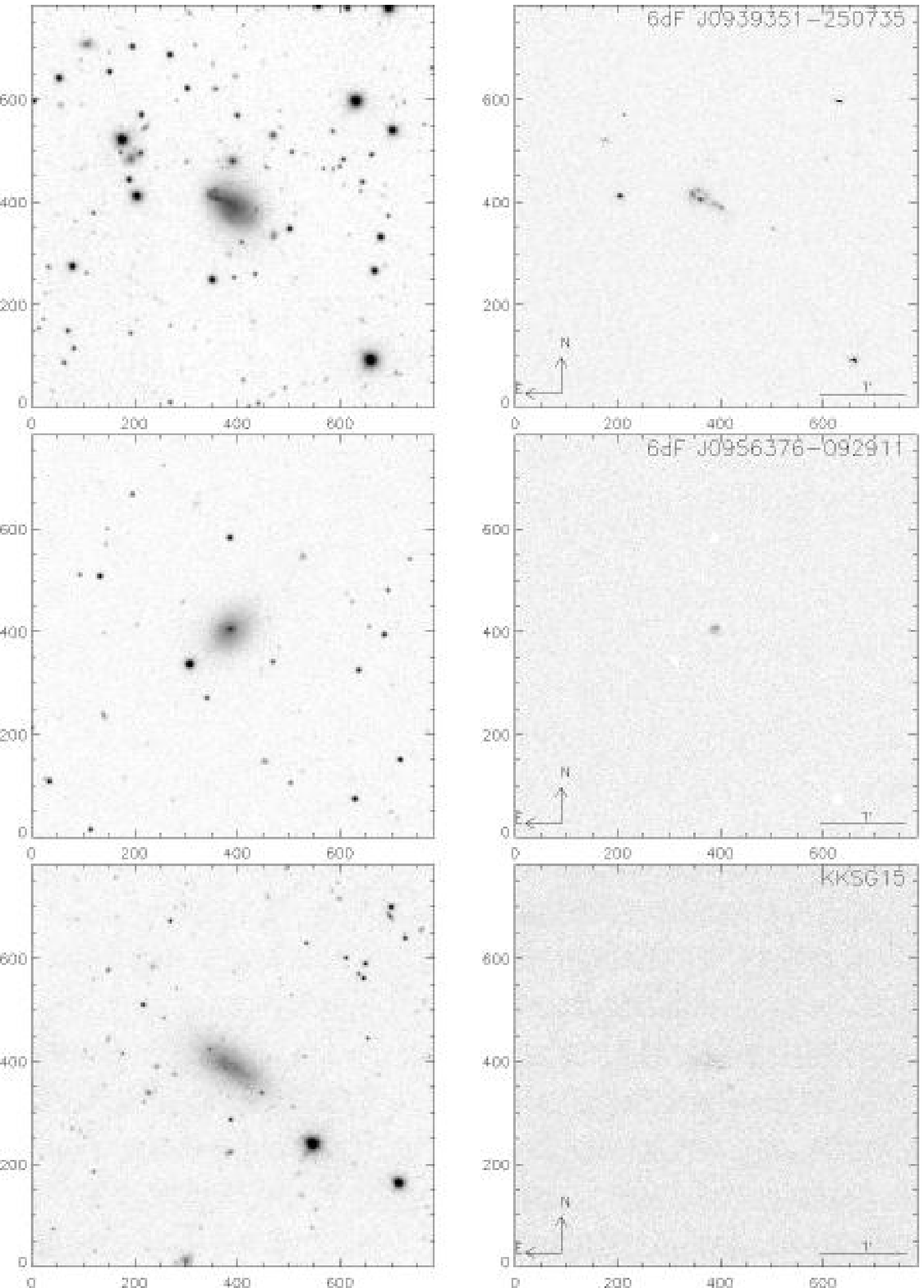}}
\vspace{22cm}
\caption{Continued}
\end{figure}

  \begin{figure}
\setcounter{figure}{0}
 \vbox{\includegraphics{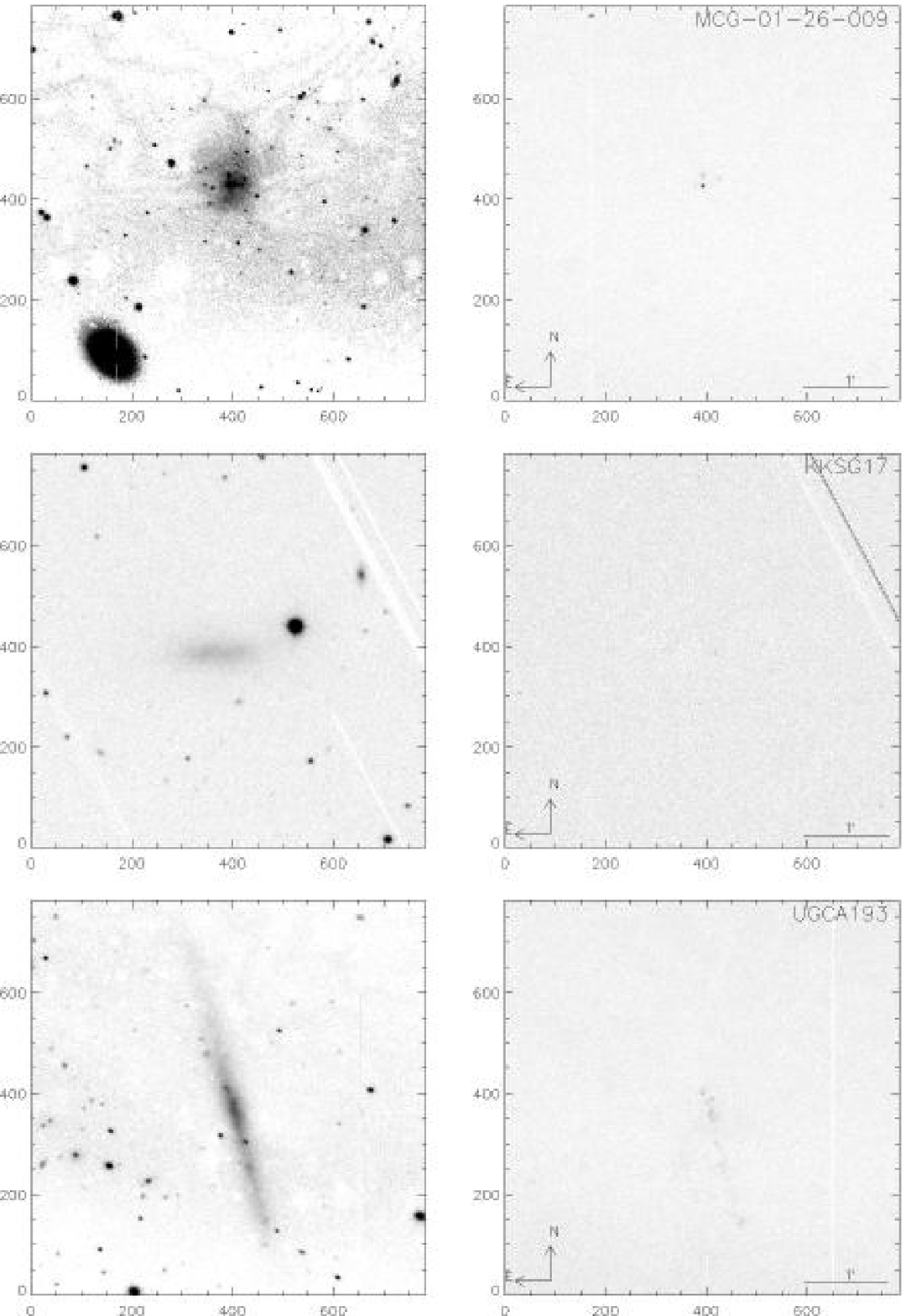}}
\vspace{22cm}
\caption{Continued}
\end{figure}

  \begin{figure}
\setcounter{figure}{0}
 \vbox{\includegraphics{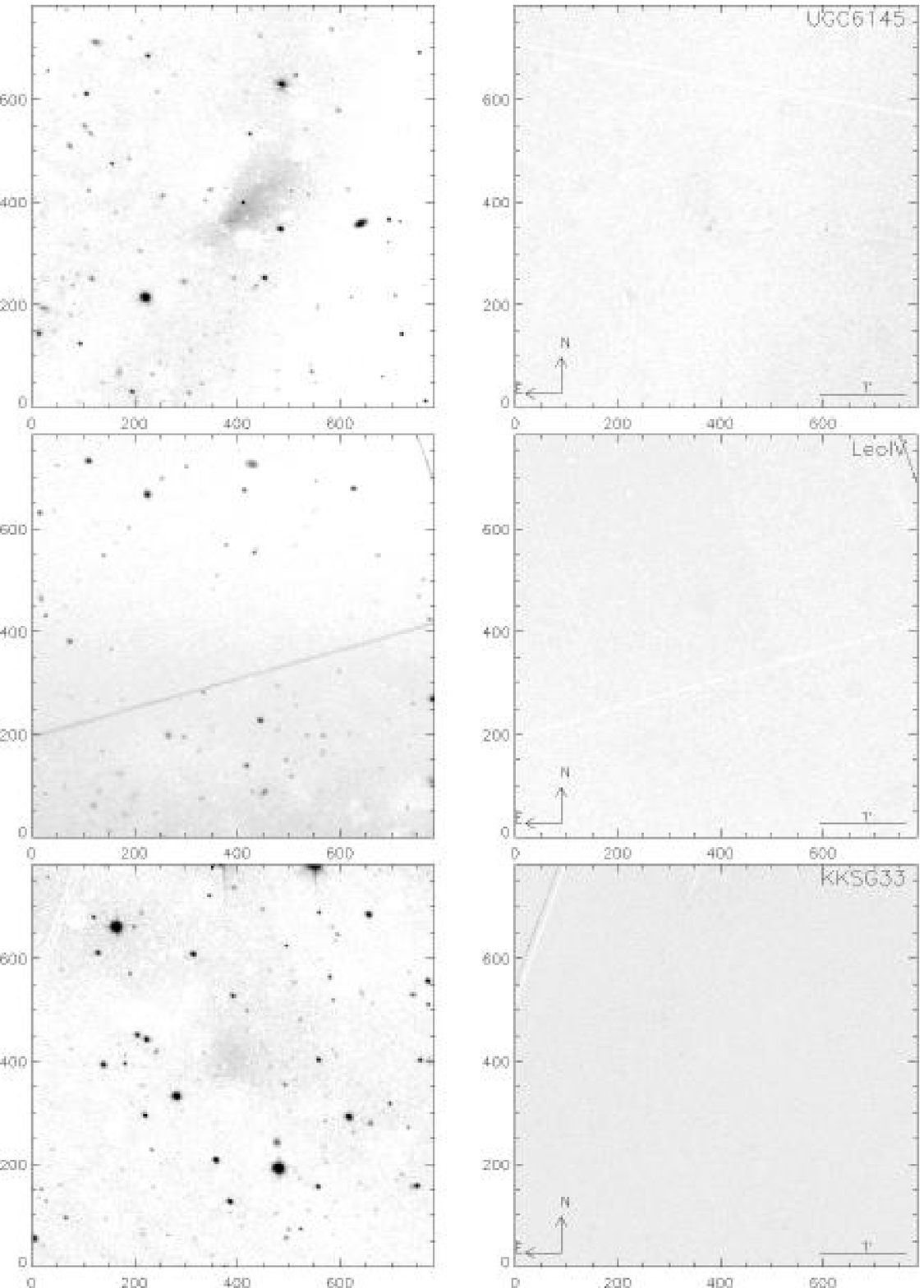}}
\vspace{22cm}
\caption{Continued}
\end{figure}

  \begin{figure}
\setcounter{figure}{0}
 \vbox{\includegraphics{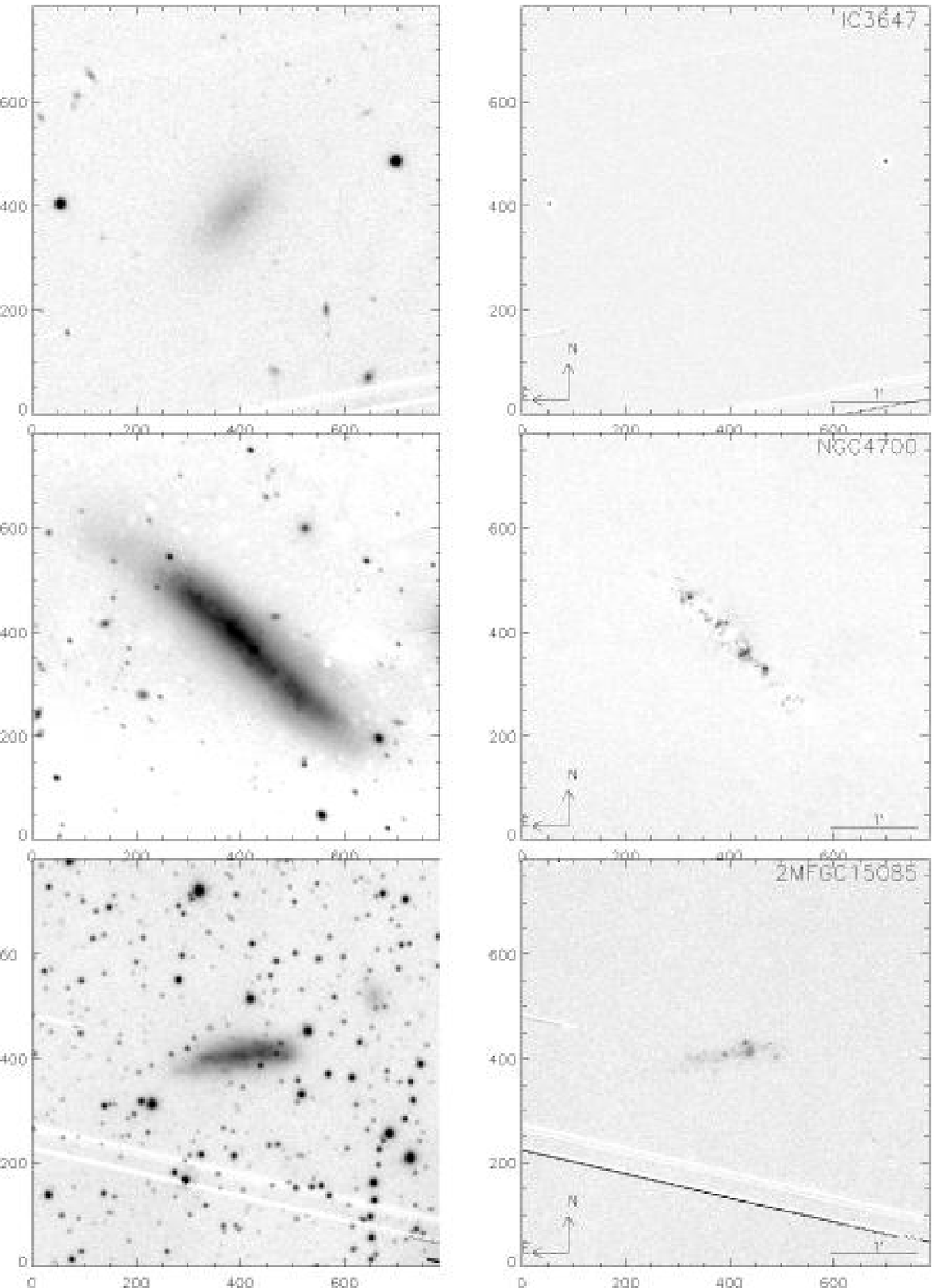}}
\vspace{22cm}
\caption{Continued}
\end{figure}

\begin{figure}
 \vbox{\includegraphics{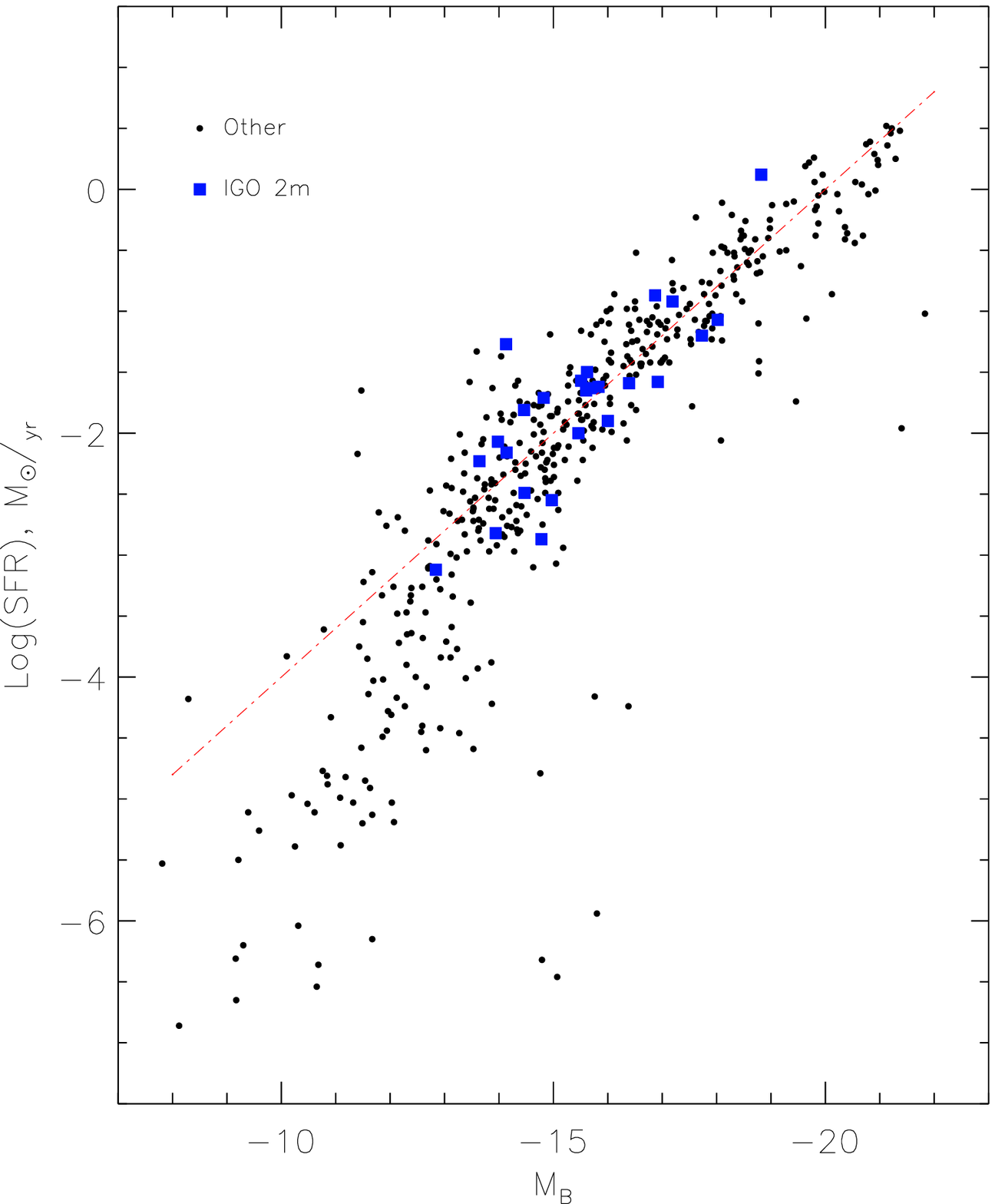}}
\vspace{22cm}
\caption{The distribution of  galaxies versus their absolute magnitude $M_B$
       and SFR.}
\end{figure}

  \begin{figure}
 \vbox{\includegraphics{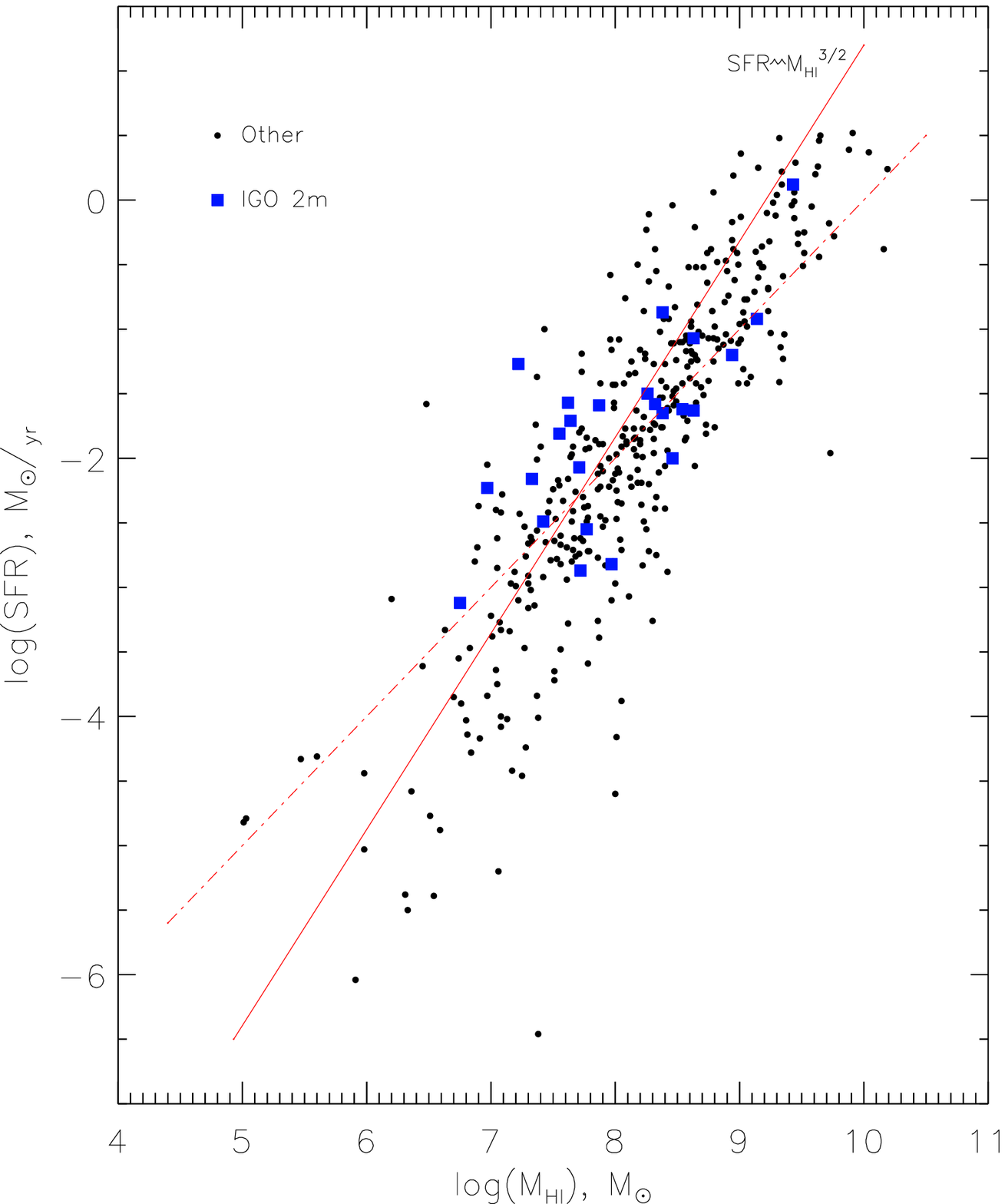}}
\vspace{22cm}
\caption{The global SFR of galaxies vs. their mass of neutral hydrogen $M_{HI}$.}
\end{figure}

  \begin{figure}
 \vbox{\includegraphics{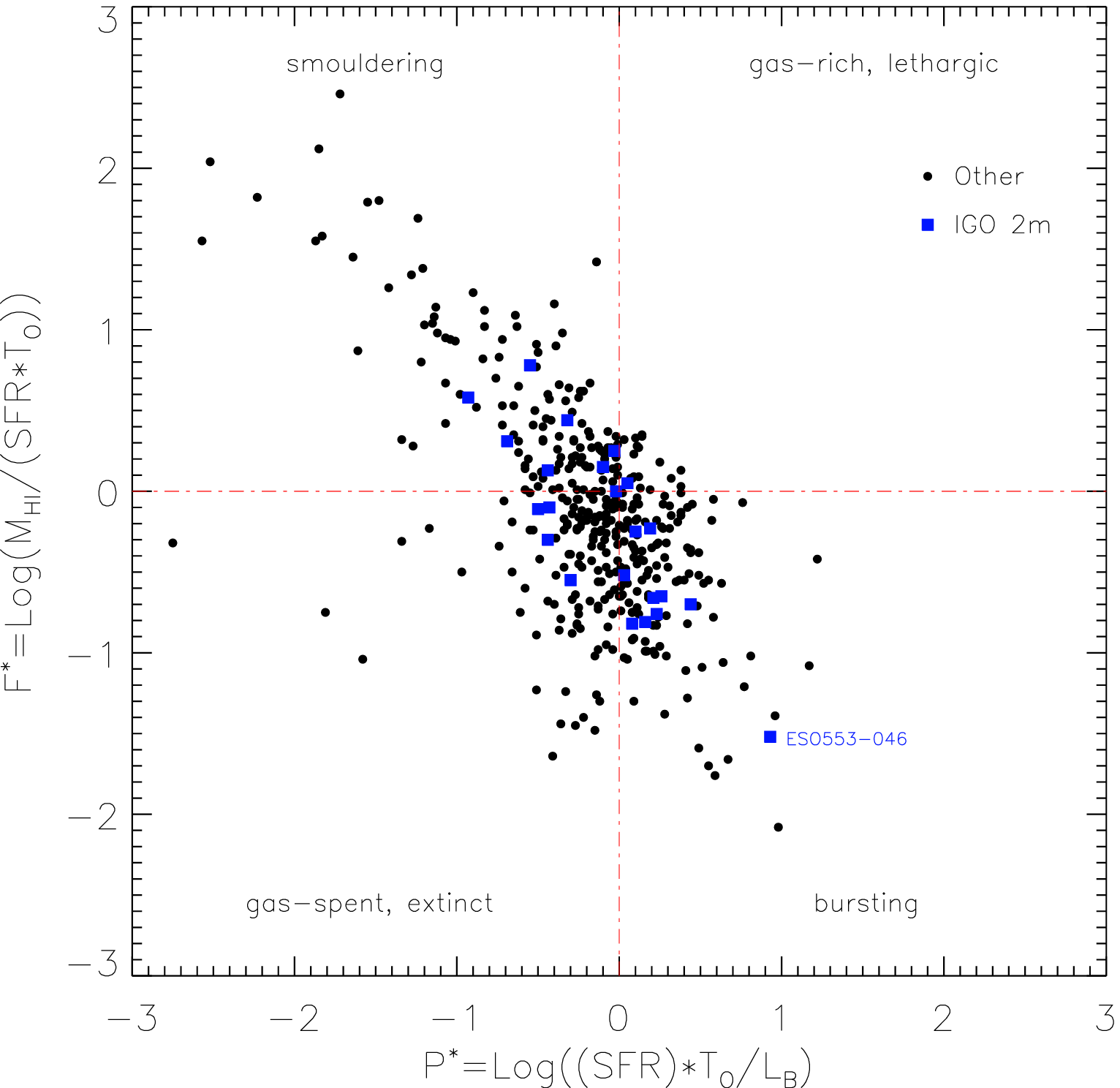}}
\vspace{22cm}
\caption{The diagnostic ``past-future'' diagram for the observed galaxies.}
\end{figure}
\end{document}